\documentclass[english,12pt,aps,prd,a4paper,preprintnumbers,floatfix,nofootinbib,showpacs,superscriptaddress,notitlepage]{revtex4-1}

\pdfoutput=1
\usepackage{subcaption}
\usepackage{graphicx}
\usepackage{rotate}
\usepackage[utf8]{inputenc}
\usepackage{amsmath}
\usepackage{amssymb}
\usepackage{float}
\usepackage{comment}
\usepackage{soul}
\usepackage[usenames,dvipsnames]{color}
\usepackage[colorinlistoftodos]{todonotes}
\usepackage[colorlinks=true,citecolor=blue,urlcolor=blue, pdfborder={0 0 0}]{hyperref}
\usepackage[normalem]{ulem}
\newcommand {\ignore}[1]{}
\usepackage{multirow}
\definecolor{darkred}{rgb}{0.6,0,0}

\newcommand{\rd}[1]{{\color{red}  #1}}

\usepackage[english]{babel}
\usepackage{blindtext}

\definecolor{midnightblue}{RGB}{25,25,112}
\definecolor{brown}{rgb}{0.59, 0.29, 0.0}

\newcommand{\AddrIISERchm}{Department of Chemistry,\\
Indian Institute of Science Education and Research (IISER) Bhopal \\
Bhopal Bypass Road, Bhauri, Bhopal 462066 INDIA}

\newcommand{\AddrIISERphy}{Department of Physics,\\
Indian Institute of Science Education and Research (IISER) Bhopal \\
Bhopal Bypass Road, Bhauri, Bhopal 462066 INDIA}
\begin{document}
\title{Work distribution of a colloid in an elongational flow field and under Ornstein-Uhlenbeck noise}
\author{Debasish Saha}
\affiliation{\AddrIISERphy}
\author{Rati Sharma}\email{rati@iiserb.ac.in}
\affiliation{\AddrIISERchm}
\begin{abstract}
\vspace{1cm}
\noindent
The study of thermodynamic properties of microscopic systems, such as a colloid in a fluid, has been of great interest to researchers since the discovery of the fluctuation theorem and associated laws of stochastic thermodynamics. However, most of these studies confine themselves to systems where effective fluctuations acting on the colloid are in the form of delta-correlated Gaussian white noise (GWN). In this study, instead, we look into the work distribution function when a colloid trapped in a harmonic potential moves from one position to another in a fluid medium with an elongational flow field where the effective fluctuations are given by the Ornstein-Uhlenbeck (OU) noise, a type of coloured noise. We use path integrals to calculate this distribution function and compare and contrast its properties to the case with GWN. We find that the work distribution function turns out to be non-Gaussian as a result of the elongational flow field, but continues to obey the fluctuation theorem in both types of noise. Further, we also look into the effects of the various system parameters on the behaviour of work fluctuations and find that although the distribution tends to broaden with increasing noise intensity, increased correlation in fluctuations acts to oppose this effect. Additionally, the system is found to consume heat from the surroundings at early times and dissipate it into the media at later times. This study, therefore, is a step towards gaining a better understanding of the thermodynamic properties of colloidal systems under non-linear complex flows that also display correlated fluctuations.
\end{abstract}
\maketitle
\section{Introduction}
\label{sec:intro}
\noindent
Thermodynamic properties of microscopic systems display markedly different behaviours compared to those exhibited by macroscopic systems. In particular, the principles that govern classical thermodynamics for macroscopic systems are often violated in the microscopic limit. For microscopic systems that are away from equilibrium, these violations appear as broad distributions of thermodynamic quantities such as entropy, heat and work \cite{chatterjee2010, sharma2011, ciliberto2013, ghosal2016, pagare2019, paraguassu2021, paraguassu2022}. The corresponding distributions have been observed in several experimental studies \cite{evans1993, wang2002, carberry2004, carberry2004_kawasaki, feitosa2004, ciliberto2004, wang2005, schuler2005, hayashi2007} and also match those calculated using the principles of non-equilibrium statistical mechanics \cite{imparato2007, ciliberto2013}. Additionally, these distribution functions are now known to follow a certain universal principle, called the fluctuation theorem (FT) \cite{evans1993, wang2002}.
\\~\\
The fluctuation theorem \cite{sevick2008} is an important result in the field of stochastic thermodynamics that provides insight into the distributions of thermodynamic quantities of systems that are in a non-equilibrium state. The FT was first proposed in 1993 \cite{evans1993}, in which Evans, Cohen and Morriss showed that there is a finite probability of entropy being consumed particularly when the system is far from equilibrium, leading to a violation of the second law of thermodynamics. Mathematically, it states that the ratio between the probability of entropy production to that of entropy consumption varies as $\exp{(\beta S)}$, where $\beta = 1/k_{B}T^{\prime}$ is the Boltzmann factor and $T^{\prime}$ is the temperature of the surrounding heat bath. This mathematical expression was later found to be true not just for entropy, but for other thermodynamic quantities as well, such as work and heat. The principle of Fluctuation theorem was experimentally verified for the first time in 2002 in which a plastic bead was trapped by an optical tweezer \cite{ashkin1997} and set to move around in a solution \cite{wang2002}. Following this, over the years, multiple other experiments have validated FT for various systems, such as systems of optically trapped colloidal beads, granular systems, systems in turbulent flows, nano-scale systems, material sciences, and also in many biological processes such as protein folding, chemical kinetics, gene regulations and RNA folding \cite{carberry2004, carberry2004_kawasaki, feitosa2004, ciliberto2004,  wang2005, schuler2005, hayashi2007, alemany2010, wong2018, baird2003, collin2005, hayashi2010, hayashi2018, park2018, sharma2012force}. Another area where stochastic thermodynamics and FT are studied extensively is in the systems of active baths in which the thermodynamics of passive tracers or colloids are studied under the influence of random collisions with active components \cite{dabelow2019, soni2023, hayashi2007}. Here, in this work, we focus on the stochastic thermodynamics of a colloidal system that is under the influence of correlated fluctuations which was earlier found to have a significant influence on different physical, chemical, and biological processes \cite{mandelbrot1968, kou2004, chaudhury2006, batra2021, batra2022}.
\\~\\
Recent advances in the study of active baths using the principles of non-equilibrium statistical mechanics have gained significant attention in the past few years \cite{wu2000, um2019, bonilla2019active, paneru2022, park2020, granek2022}. Active baths are composed of self-propelling units that can convert energy from the surroundings to directed motions. Because of this reason, a bath consisting of active components is always present in a non-equilibrium state. Understanding the behaviour of such systems has become the central interest of both theoreticians and experimentalists from very diverse research backgrounds. In order to model the dynamics of a passive tracer in such an active system, an active random force is considered in the equation of motion in addition to the Gaussian white noise (GWN) arising due to the thermal fluctuations in the system. The active counterpart of the random force appears from the random collisions of the passive tracer with the surrounding active components in the system. The active force is usually modelled using the Gaussian coloured noise, also known as the {\it Ornstein-Uhlenbeck} (OU) noise which has exponential correlations in time with a characteristic time limit ($\tau$) \cite{uhlenbeck1930}. The incorporation of this OU noise then enables a study of the time evolution of thermodynamic observables of such systems away from equilibrium  \cite{wu2000, um2019, bonilla2019active, paneru2022, park2020, granek2022}. Further, in the study of active baths, the contribution of thermal fluctuations can be ignored when the velocities of the active components are very high \cite{wu2000, angelani2010, knevzevic2020}.
\\~\\
Although significant efforts have gone towards the study of stochastic thermodynamics in colloidal systems, its understanding in the presence of varying background fluid flows and velocities is still lacking. Earlier studies often considered the background medium of these systems to be either at rest or having a uniform velocity. The study of the motion of colloids in the presence of a constant flow of the surrounding medium \cite{imparato2007} or that of a charged Brownian oscillator in an external electric field \cite{singh2008} are a couple such examples. But, in practice, whether it is a particle moving in an air medium or a cellular entity moving around in the cytoplasm, the nature of the motion of background media has a significant effect on the dynamics of the tracer particle. The media in such real systems in which colloids move around are found to exhibit motions that are non-uniform in nature and that keep the systems away from equilibrium. Also in the case of active baths, the motion of active components generates disorder in the media which in turn affects the motion of the passive tracers. Therefore, for the purposes of generality, it becomes important to study the statistical behaviours of such systems. But, unfortunately, the thermodynamics of such systems are very poorly studied because of the complications that arise from the complex flow patterns present in the surrounding media, rendering the flow anisotropic. Further, although the types and complexities of non-linearity can vary from one system to another, in practice, it is preferable to study some particular types of non-linear gradient flows such as one or a combination of shear, rotation and elongation, to get insights into the dynamics of a particle under such conditions. Some of the past research works have focused on the dynamics and thermodynamics of colloids in various types of non-linear flows, including shear and elongational flow \cite{perkins1997, smith1999, chatterjee2010, sharma2011, sharma2012, sharma2012JCP, latinwo2013, latinwo2014, ghosal2016, pagare2019, ghosal2019, pelargonio2023generalized}. In a few of these works, the probability distributions of different observables were also evaluated which were found to satisfy the fluctuation theorem \cite{sharma2011, chatterjee2010, ghosal2016, pagare2019}. Most of these studies, however, were carried out considering delta-correlated noise, also known as Gaussian white noise (GWN) arising from thermal fluctuations that are intrinsic to the system \cite{zwanzig2001, chatterjee2010, sharma2011, sharma2013, sharma2012, singh2008, balakrishnan2008}. But, the thermodynamics of such systems in the presence of OU noise (or coloured noise) which is often considered as an external noise \cite{wu2000, park2020, ye2020, roberts2015, paneru2022, sharma2022}, still remains unexplored.
\\~\\
In this article, we use the path integral technique \cite{chaichian2001, feynman1965} to evaluate the work distribution function of a colloidal particle moving from one position to another in a fluid medium that is exhibiting a particular non-linear flow and where the colloid is itself under the influence of an external harmonic potential. We define our system and the corresponding equation of motion of a colloidal particle in elongational flow and in the presence of external harmonic potential through the overdamped Langevin equation and then go on to calculate the work done by the colloid in moving from one position to another. We introduce the elongational fluid flow in this system as it is a typical example of a non-linear flow of a fluid in which the $x$ and $y$ components of the velocity field ($\vec{v} (\vec{r})$) are coupled with each other while the $z$ component of velocity remains independent of other components. Because of this coupling behaviour, the work done in moving the colloid from one position to another is a non-linear function of its time-dependent coordinates and the distribution of work done (which is the primary goal of our study) becomes asymmetric. We also compare this with the work distribution for constant background flow by modifying the velocity field such that the flow rate ($\dot{\gamma}$) is set to zero in the expression of $\vec{v} (\vec{r})$. For both types of background flow, we study the dynamics of the colloid by considering different types of noise such as white noise and coloured noise to examine how the work distributions change when internal and external noise are considered, respectively.
\\~\\
The rest of the manuscript is organized as follows. The system under consideration in this study and the amount of work done due to the motion of the colloid is described in Section \ref{sec:system_description}. The calculation of the probability distribution function for the final positions of the colloid in an elongational flow field using OU or coloured noise is shown in Section \ref{sec:cond_prob_ou}. A similar calculation for delta-correlated or white noise is given in Appendix \ref{app:pdf_delta}. The calculation of the work distribution function and the corresponding results and discussion are presented in Section \ref{sec:results}.
\section{Motion of a colloid in an elongational flow field}
\label{sec:system_description}
\noindent
Consider a colloidal particle that is under the influence of an external harmonic potential and is moving in a fluid medium with an elongational flow. The dynamics of this colloid is highly affected by the noise present in the system. Let, $\vec{r}(t)$ be the position of the colloid at any time $t$. The motion of the colloid in such a system can be described by an overdamped Langevin equation \cite{langevin1908, lemons1997}, as follows,
\begin{equation}
\label{eq:langevin}
    \zeta \dot{\vec{r}}-\zeta\vec{v}(\vec{r})+\frac{\partial U(\vec{r})}{\partial\vec{r}} = \vec{\eta} (t)
\end{equation}
where $\zeta$ is the friction coefficient. The velocity profile of the solvent medium is given by $\vec{v} (\vec{r}) = \vec{v}_{0} + \dot{\gamma} \kappa \vec{r}$, where $\vec{v}_{0} = v_{0} (\hat{i}+\hat{j}+\hat{k})$ is the constant background solvent velocity, $\dot{\gamma}$ is the flow rate and $\kappa=
\Big(\begin{smallmatrix}
  0 & 1 & 0 \\
  \alpha & 0 & 0 \\
  0 & 0 & 0
\end{smallmatrix}\Big)$ is the velocity gradient tensor which is responsible for the non-linear flow of the medium and coupling between different velocity components. The value of $\alpha$ varies from $-1$ to $+1$. $\alpha = -1$ corresponds to pure rotation, 0 corresponds to shear flow and $+1$ corresponds to elongational flow. In our calculations, we have used $\alpha = 1$ which is the case for elongational type of flows. The external harmonic potential is given by $U(\vec{r}) = k r^{2}/2$. $\vec{\eta} (t)$ is the random force (or noise) acting on the colloid. This noise, can either be uncorrelated, as is the case for thermal fluctuations, or correlated, as is the case for active noise. Here, we consider the noise to be in the form of the OU process \cite{uhlenbeck1930}, wherein, the noise auto-correlation function decays exponentially over time with a characteristic time constant of $\tau$. The OU process is a stochastic process whose dynamics can be represented in terms of the following stochastic differential equation: $\dot{\eta}(t) = -\eta(t)/\tau + \sqrt{D}\theta(t)/\tau$. Here $\theta (t)$ is a Gaussian white noise of zero mean and delta-correlated autocorrelation function. $D$ represents the strength of the noise. The OU noise, therefore, has the following statistical properties:
\begin{subequations}
\label{eq:ou_noise_prop}
    \begin{equation}
        \big\langle \vec{\eta}_{i} (t) \big\rangle = 0
    \end{equation}
    \begin{equation}
        \big\langle \vec{\eta}_{i} (t) \vec{\eta}_{j} (t^{\prime}) \big\rangle = \frac{D}{\tau} \delta_{ij} \exp\bigg(\frac{-|t-t^{\prime}|}{\tau}\bigg)
    \end{equation}
\end{subequations}
with higher $D$ corresponding to larger fluctuations about the mean. 
\\~\\
The colloid moves in the fluid from one position to another during a time interval of $T$ and during the event it performs some work due to its motion. Since the process is stochastic, the trajectories between the initial and final points for the given interval are different for each sample. As a result, the work done, which is a path-dependent function, also varies from sample to sample. This ultimately leads to a distribution of work for a specific set of system conditions. This work done by the colloid following a particular trajectory can, in general, be calculated as,
\begin{equation}
\label{eq:work}
    W_{T}=\int_{0}^{T} \vec{v}(\vec{r})\cdot \nabla U(\vec{r}) ~dt
\end{equation}
Here, $\vec{v}(\vec{r})$ is the velocity of the solvent and $U(\vec{r})$ is the external harmonic potential as mentioned earlier. For the evaluation of the work distribution under the influence of the OU noise, we consider two special cases. The first case is for the constant background flow which is obtained by setting $\dot{\gamma}=0$ in the expression of $\vec{v}(\vec{r})$, whereas the second case is for elongational flow with nonzero $\dot{\gamma}$. We discuss these two cases in detail in Section \ref{sec:results}. We next compute the conditional probability distribution of the position of the colloidal particle before moving on to the calculation of the work distribution function. In the following calculations, we have used different values of $T$ to show the system behaviour at different times. The probability distributions for the final positions of the colloid in different cases are shown at two different times ($T = 1$ and $10$). Further, in the plots to illustrate the fluctuation theorem, we have used different values of $T$ (upto 2.7 for OU noise in elongational flow) to show the dynamical changes in the probability distribution of positive and negative work done along a trajectory. Also, in the study of the dynamical change of distribution properties, we used a range of time scales depending on the choice of various parameters such as friction coefficient, relaxation time, etc., for different cases. For example, in the plots showing the mean work and standard deviation, $T$ is varied between 0 and 10 for constant flow and between 0 and 14 for elongational flow. Similarly, in the plots showing the skewness parameter, $T$ is varied between 0 and 6 for white noise and between 0 and 15 for coloured noise. Numerical computations exhibit divergence errors beyond these values of $T$ and therefore have not been plotted. However, all the interesting characteristics of the quantities can already be seen for the plotted ranges of time.

\section{Conditional Probability distribution for the final position of the colloid}
\label{sec:cond_prob_ou}

\noindent
We now focus on computing the probable distance the colloidal particle can travel in a time $T$ given that it was at position $\vec{r}_0$ at the initial time. The resulting conditional probability can then be used to compute the distribution function for the work performed. Since the OU process \cite{uhlenbeck1930} is Gaussian distributed, the probability distribution of the colloid following a particular trajectory during a time interval $T$ can be obtained from \cite{mckane1990, bray1990, luckock1990}
\begin{equation}
\label{eq:prob_dist_ou}
    P[\vec{\eta}] \propto \exp\bigg\{-\frac{1}{4D}\int_{0}^{T}dt\Big[\vec{\eta} (t)^{T}\vec{\eta}(t)+\tau^{2}\dot{\vec{\eta}}(t)^{T}\dot{\vec{\eta}}(t)\Big]\bigg\}
\end{equation}
Using $\vec{\eta}$ and its first order time derivative ($\dot{\vec{\eta}}$) from Eq. \ref{eq:langevin} in Eq. \ref{eq:prob_dist_ou}, we get 
\begin{equation}
\begin{split}
\label{eq:prob_dist_positions}
    P[x,y,z] \propto J[x,y,z] \exp\bigg\{-\frac{1}{4D}\int_{0}^{T}dt\Big[\tau^{2}\zeta^{2}(\ddot{x}^{2}+\ddot{y}^{2}+\ddot{z}^{2})+(\zeta^{2}+\tau^{2}k^{2}+\tau^{2}\zeta^{2}\dot{\gamma}^{2})(\dot{x}^{2}+\dot{y}^{2}) \\
    +(\zeta^{2}+\tau^{2}k^{2})\dot{z}^{2}-2\zeta^{2}v_{0}(\dot{x}+\dot{y}+\dot{z})+2\zeta k(x\dot{x}+y\dot{y}+z\dot{z})+2\zeta k\tau^{2}(\dot{x}\ddot{x}+\dot{y}\ddot{y}+\dot{z}\ddot{z}) \\ -2\zeta^{2}\dot{\gamma}(\dot{x}y+x\dot{y})-2\zeta^{2}\dot{\gamma}\tau^{2}(\dot{x}\ddot{y}+\ddot{x}\dot{y})-4\zeta\dot{\gamma}k\tau^{2}\dot{x}\dot{y}-4k\zeta\dot{\gamma}xy+(k^{2}+\zeta^{2}\dot{\gamma}^{2})(x^{2}+y^{2}) \\
    +k^{2}z^{2}+(2\zeta^{2}\dot{\gamma}v_{0}-2\zeta kv_{0})(x+y)-2\zeta kv_{0}z+3\zeta^{2}v_{0}^{2} \Big]\bigg\}
\end{split}
\end{equation}
where, $J[x,y,z]$ is the Jacobian for the change of variable from $\vec{\eta}$ to $\vec{r}$ \cite{sharma2011, chatterjee2010} whose calculation is shown in Appendix \ref{app:jacobian}. The conditional probability density, $P(x_{f},y_{f},z_{f},T|x_{0},y_{0},z_{0})$, of finding the particle at ($x_{f},y_{f},z_{f}$) after time $T$ given that the particle started moving from ($x_{0},y_{0},z_{0}$) at $t=0$, can be expressed as
\begin{equation}
\begin{split}
\label{eq:cond_prob_ou}
    P(x_{f},y_{f},z_{f},T | x_{0},y_{0},z_{0}) \propto ~& e^{3kT/2\zeta}e^{-\frac{\zeta k}{4D}\big[(x_{f}^{2}+y_{f}^{2}+z_{f}^{2}-x_{0}^{2}-y_{0}^{2}-z_{0}^{2})+\tau^{2}(v_{x_f}^{2}+v_{y_f}^{2}+v_{z_f}^{2}-v_{x_0}^{2}-v_{y_0}^{2}-v_{z_0}^{2})\big]} \\
    &\times\int_{x(0)=x_{0}}^{x(T)=x_{f}} \mathcal{D}[x] \int_{y(0)=y_{0}}^{y(T)=y_{f}} \mathcal{D}[y] \int_{z(0)=z_{0}}^{z(T)=z_{f}} \mathcal{D}[z] ~e^{-S[x,y,z]}
\end{split}
\end{equation}
where $\mathcal{D}[x]$, $\mathcal{D}[y]$ and $\mathcal{D}[z]$ represent the path integrals over $x$, $y$ and $z$ between the end points ($x_{0},y_{0},z_{0}$) and ($x_{f},y_{f},z_{f}$), and $S[x,y,z]$ represents the action during the time interval of $T$,  defined as
\begin{equation}
\label{eq:action_ou}
    S[x,y,z] = \int_{0}^{T} dt ~\mathcal{L}(x,y,z,\dot{x},\dot{y},\dot{z},\ddot{x},\ddot{y},\ddot{z},t)
\end{equation}
Here, $\mathcal{L}$ represents the Lagrangian of the system given by,
\begin{equation}
\begin{split}
\label{eq:lagrangian_ou}
    \mathcal{L}(x,y,z,\dot{x},\dot{y},\dot{z},\ddot{x},\ddot{y},\ddot{z},t) = \frac{1}{4D} \Big[ \tau^{2}\zeta^{2}(\ddot{x}^{2} + \ddot{y}^{2} + \ddot{z}^{2}) +(\zeta^{2} + \tau^{2}k^{2} + \tau^{2}\zeta^{2}\dot{\gamma}^{2}) (\dot{x}^{2}+\dot{y}^{2}) \\
    +(\zeta^{2}+\tau^{2}k^{2})\dot{z}^{2}-2\zeta^{2}v_{0}(\dot{x}+\dot{y}+\dot{z})-2\zeta^{2}\dot{\gamma}(\dot{x}y+x\dot{y})-2\zeta^{2}\dot{\gamma}\tau^{2}(\dot{x}\ddot{y}+\ddot{x}\dot{y})-4\zeta\dot{\gamma}k\tau^{2}\dot{x}\dot{y} \\ -4k\zeta\dot{\gamma}xy+(k^{2}+\zeta^{2}\dot{\gamma}^{2})(x^{2}+y^{2})+k^{2}z^{2}+(2\zeta^{2}\dot{\gamma}v_{0}-2\zeta kv_{0})(x+y)-2\zeta kv_{0}z+3\zeta^{2}v_{0}^{2} \Big]
\end{split}
\end{equation}
\\~\\
Eq. \ref{eq:cond_prob_ou} represents the path integral for the particle moving from the initial to the final position that can be solved using Feynman’s variational technique \cite{chaichian2001, feynman1965}. The motion of the colloid in such a system is highly stochastic and there can be an infinitely large number of trajectories between the initial and final points. The most probable trajectory along which the action is minimum can therefore be found using the Euler-Lagrange equation of motion, given by
\begin{equation}
\label{eq:euler_lagrange_eom}
    \frac{\partial \mathcal{L}}{\partial r_{i}} - \frac{d}{dt}\bigg(\frac{\partial \mathcal{L}}{\partial \dot{r}_{i}}\bigg) + \frac{d^2}{dt^2}\bigg(\frac{\partial \mathcal{L}}{\partial \ddot{r}_{i}}\bigg)=0
\end{equation}
where the index $i=1,2,3$ corresponds to $x$, $y$ and $z$ components. Using the Lagrangian in Eq. \ref{eq:euler_lagrange_eom}, the equation of motion of the colloid becomes
\begin{subequations}
\begin{equation}
\label{eq:eom_ou}
    \ddddot{\vec{r}} + M\ddot{\vec{r}} + N\vec{r} + P\vec{I} = 0
\end{equation}
\text{where}
\begin{equation}
    M=\begin{pmatrix}
    -\alpha_{1} & \alpha_{2} & 0 \\
    \alpha_{2} & -\alpha_{1} & 0 \\
    0 & 0 & -\beta_{1} \\
    \end{pmatrix}; 
    N=\begin{pmatrix}
    \alpha_{3} & -\alpha_{4} & 0 \\
    -\alpha_{4} & \alpha_{3} & 0 \\
    0 & 0 & -\beta_{2} \\
    \end{pmatrix}; 
    P=\begin{pmatrix}
    \alpha_{5} \\
    \alpha_{5} \\
    -\beta_{3} \\
\end{pmatrix}
\end{equation}
\end{subequations}
and $\vec{I}$ is the $3\times3$ identity matrix. Here, $\alpha_{1}=\dot{\gamma}^{2}+\frac{1}{\tau^{2}}+\frac{k^{2}}{\zeta^{2}}$, $\alpha_{2}=\frac{2k\dot{\gamma}}{\zeta}$, $\alpha_{3}=\frac{k^{2}}{\tau^{2}\zeta^{2}}+\frac{\dot{\gamma}^{2}}{\tau^{2}}$, $\alpha_{4}=\frac{2k\dot{\gamma}}{\zeta\tau^{2}}$, $\alpha_{5}=\frac{\dot{\gamma}v_{0}}{\tau^{2}}-\frac{kv_{0}}{\zeta\tau^{2}}$, $\beta_{1}=\frac{1}{\tau^{2}}+\frac{k^{2}}{\zeta^{2}}$, $\beta_{2}=\frac{k^2}{\tau^2\zeta^2}$ and $\beta_{3}=\frac{kv_{0}}{\zeta\tau^{2}}$.
\begin{figure}
    \centering
    \includegraphics[width=15cm]{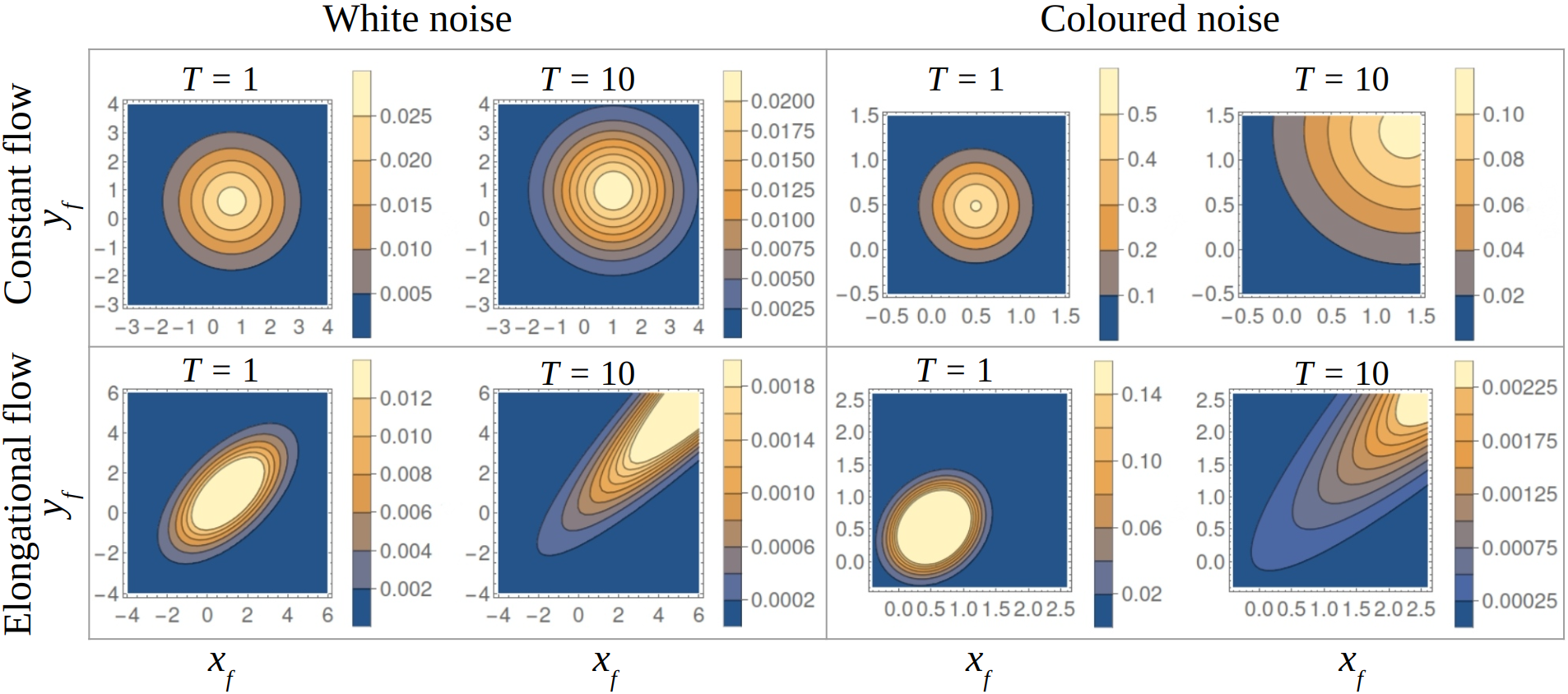}
    \caption{Distribution for the final positions of the colloid (at $T=1$ and 10) in a fluid having constant flow (upper row) and elongational flow (lower row) in the presence of Gaussian white noise (left) and OU noise (right). Distributions are calculated for $\dot{\gamma}=1$ in the case of elongational flow, whereas $\dot{\gamma}$ is taken to be zero for constant flow field. Friction coefficient ($\zeta$), stiffness constant ($k$), background velocity of the fluid ($v_{0}$), noise strength ($D$) and relaxation time ($\tau$) were fixed to unity during these calculations. Colourbar represents the value of the conditional probability density.}
    \label{fig:pdf_position}
\end{figure}
\\~\\
The $x$ and $y$ components of Eq. \ref{eq:eom_ou} are fourth-order coupled differential equations which are difficult to solve analytically. However, in the limiting case of constant background flow that can be obtained from the general case of colloid in an elongational flow medium by setting $\dot{\gamma}=0$, equations of motions along individual components are independent of each other and, therefore, are possible to solve analytically. We solve for $x(t)$, $y(t)$ and $z(t)$ numerically using Mathematica \cite{wolfram2022} by first setting the parameters of Eq. \ref{eq:eom_ou} to specific constant values. We set $\tau$, $\zeta$, $k$ and $D$ to unity (in appropriate units) to avoid extremely large solutions. We consider $v_{0}$ and $\dot{\gamma}$ to be unity as well. Solutions of $x(t)$ and $y(t)$ contain eight constants that can be evaluated by using the boundary conditions, which are $x(0)=x_{0}$, $x(T)=x_{f}$, $\dot{x}(0)=v_{x_{0}}$, $\dot{x}(T)=v_{x_{f}}$ and $y(0)=y_{0}$, $y(T)=y_{f}$, $\dot{y}(0)=v_{y_{0}}$, $\dot{y}(T)=v_{y_{f}}$. Solution for the $z$-component is comparatively easy since the motion of the particle along the $z$-direction is independent of the other components, and therefore can be solved using the boundary conditions $z(0)=z_{0}$, $z(T)=z_{f}$, $\dot{z}(0)=v_{z_{0}}$ and $\dot{z}(T)=v_{z_{f}}$. For further simplification, we consider that the motion of the colloid starts from the origin at $t=0$ with zero initial velocity. We took the final velocity of the colloid to be unity as well along each direction. The action was then calculated using all these values of parameters within a time scale of $0$ to $T$. Taking all these constants, the final form of the normalized conditional PDF at some arbitrary time $T$ is
\begin{equation}
\label{eq:final_pdf_ou}
\begin{split}
    P_{N}(x_{f},y_{f},z_{f},T | x_{0},y_{0},z_{0}) = \exp\Big[A_{1}+A_{2}(x_{f}^{2}+y_{f}^{2})+A_{3} (x_{f}+y_{f})+A_{4} x_{f} y_{f}+A_{5}z_{f}^{2}+A_{6}z_{f})\Big]
\end{split}
\end{equation}
where $A_{i}$'s are numerical constants depending upon the particular choice of parameters mentioned above. A similar calculation of the conditional probability distribution for the case of a colloid in a fluid flow under the influence of delta-correlated noise is provided in Appendix \ref{app:pdf_delta}. The plots for the conditional probability distributions (as obtained from Eq. \ref{eq:final_pdf_ou} and Eq. \ref{eq:final_pdf_delta}) for the final position of the colloid in the case of constant flow (upper row) and elongational flow (lower row) are shown in Fig. \ref{fig:pdf_position}. For each type of flow, we compare and contrast the distributions obtained considering white noise and coloured noise at two different final times, i.e., $T=1$ and $T=10$. This helps in comparing the shift of the distributions with time in different conditions. As seen from Fig. \ref{fig:pdf_position}, for the case of constant flow, the distributions are shifted towards the direction of the background flow field and the diffusion happens symmetrically along every direction. But, in the case of the elongational flow, distributions indeed shift along the flow field direction but the shape is elongated along the diagonal axis of the $x-y$ plane. This is expected particularly due to the nature of the flow of the surrounding medium as the flow of the medium is no longer uniform and is biased in a particular direction. As a result, the motion of the particle is more probable along the direction of flow compared to that along the other directions. One can also observe that the distributions in the case of coloured noise spread slower than that in the case of white noise for both types of flow.
\\~\\
Eq. \ref{eq:final_pdf_ou} is now further used for the calculations of work distribution in moving the colloid from the initial position to the final position which is discussed in Sec. \ref{sec:results}.
\section{Results and Discussion}
\label{sec:results}
\begin{figure*}
    \centering
    \includegraphics[width=15cm]{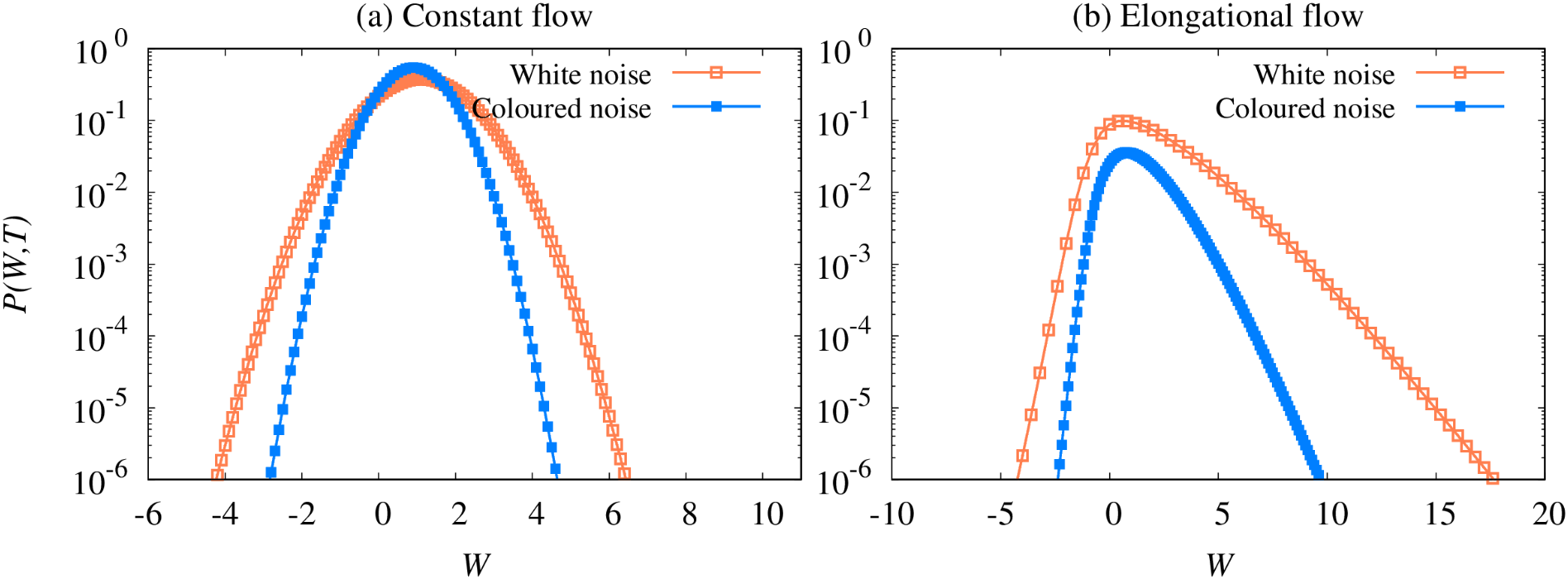}
    \caption{Work distributions of the colloid moving in (a) constant flow and (b) elongational flow considering white noise (empty squares) and coloured noise (filled squares). Distributions were calculated at a final time $T=1$ and other parameters such as friction coefficient ($\zeta$), stiffness constant ($k$), noise strength ($D$), and linear part of the background velocity ($v_0$) were fixed to unity. The relaxation time ($\tau$) in the case of OU noise was fixed at 0.1 and $\dot{\gamma}$ in the case of elongational flow was set to 0.5.}
    \label{fig:work_dist}
\end{figure*}

\noindent
Having computed and studied the dynamics of the particle through the conditional probability distribution for the final position of the particle after time $T$ (Eq. \ref{eq:final_pdf_ou} and Fig. \ref{fig:pdf_position}), we can now proceed with the calculation of the work distribution. Specifically, we calculate the distribution for work performed by the colloid during its evolution from the initial position ($x_0,y_0,z_0$) at $t=0$ to the final position ($x_f,y_f,z_f$) at any arbitrary time $T$. This distribution, given by $P(W,T)$ and representing the amount of work $W_{T}$ that is being performed in time $T$, can be expressed as,
\begin{equation}
\label{eq:work_dist_dirac_delta}
    P(W,T) = \big\langle \delta (W-W_{T}) \big\rangle
\end{equation}
The angular brackets here denote the ensemble average taken over all possible trajectories between the initial and the final positions.
\\~\\
Using the Fourier representation of the Dirac-delta function and taking the ensemble average over all possible trajectories, Eq. \ref{eq:work_dist_dirac_delta} can be re-written as
\begin{subequations}
    \begin{equation}
    \begin{split}
    \label{eq:work_dist_fourier}
        P(W,T)=e^{3kT/2\zeta}\int_{-\infty}^{\infty}d\lambda \int_{-\infty}^{\infty}dx_{0} \int_{-\infty}^{\infty}dy_{0} \int_{-\infty}^{\infty}dz_{0} \int_{-\infty}^{\infty}dx_{f} \int_{-\infty}^{\infty}dy_{f} \int_{-\infty}^{\infty}dz_{f} \\
        \times P_{0}(x_{0},y_{0},z_{0}) P_{N}(x_{f},y_{f},z_{f},T | x_{0},y_{0},z_{0}) \exp\big[i\lambda (W-W_{T})\big]
    \end{split}
    \end{equation}
\text{where,}
    \begin{equation}
    \label{eq:initial_dist}
        P_{0}(x_{0},y_{0},z_{0})=\delta(x_{0})\delta(y_{0})\delta(z_{0})
    \end{equation}
\end{subequations}
is the initial distribution of the colloid assuming that the colloid begins its motion from the origin. Substituting this initial distribution in Eq. \ref{eq:work_dist_fourier} and carrying out the integration over all the possible initial and final positions of the colloid, the characteristic function of $P(W,T)$ can be obtained as 
\begin{equation}
\label{eq:characteristic_function}
    \mathcal{C}_{W}(\lambda) = \big\langle \exp(-i\lambda W_{T})\big\rangle.
\end{equation}
Making use of this characteristic function, $\mathcal{C}_{W}(\lambda)$, the distribution for work can be calculated as
\begin{equation}
\label{eq:work_dist_final}
    P(W,T) = \int_{-\infty}^{\infty}d\lambda ~\exp(i\lambda W) ~\mathcal{C}_{W}(\lambda)
\end{equation}
Further, the moments of the work distribution function can be found analytically from the characteristic function using the formulas
\begin{equation}
    \label{eq:moments12}
    \big\langle W \big\rangle = i\frac{\partial}{\partial \lambda} \mathcal{C}_{W}(\lambda)\bigg|_{\lambda=0} \hspace{0.8cm} \text{and} \hspace{0.8cm}
    \big\langle W^{2} \big\rangle = -\frac{\partial^2}{\partial \lambda^2} \mathcal{C}_{W}(\lambda)\bigg|_{\lambda=0}
\end{equation}
\\~\\
The first moment, $\big\langle W \big\rangle$ represents the mean value of $P(W,T)$ and $\sigma= \sqrt{\langle W^2 \rangle - \langle W \rangle^2}$ gives the standard deviation of the distribution of work done. One can now calculate these properties of the work distribution function for different types of flow and in the presence of different types of noise. 
\\~\\
Work done in moving the colloid in the velocity field $\vec{v}(\vec{r}) = \vec{v}_{0} + \dot{\gamma} \kappa \vec{r}$, for a duration of time $T$ is calculated via Eq. \ref{eq:work} and is given by,
\begin{equation}
\label{eq:work_elongational}
    W_{T} = k\int_{0}^{T} \Big[v_{0}(x+y+z) + 2\dot{\gamma}xy \Big] dt
\end{equation}
We now substitute the above equation (Eq. \ref{eq:work_elongational}) into Eq. \ref{eq:work_dist_final} via Eq. \ref{eq:characteristic_function} and compute the work distribution function.
\\~\\
Specifically, for the limiting case of the constant background flow, we set the value of the flow rate ($\dot{\gamma}$) to zero which makes $W_{T}$, as given by Eq. \ref{eq:work_elongational}, a linear function of the colloid's position. The corresponding work distribution is evaluated numerically by setting the friction coefficient ($\zeta$), stiffness constant ($k$), and constant background velocity of the fluid ($v_{0}$) to unity. The relaxation time ($\tau$) is fixed to a value of 0.1. The resulting distributions for the case of delta-correlated and exponentially correlated noise at $T=1$ (in the regime where the mean work and standard deviation of the distribution rapidly increase with time) are shown in Fig. \ref{fig:work_dist}\rd{a} (shown with empty squares and filled squares, respectively). For both the cases, the distributions under the condition of constant background flow are symmetric about the mean values resembling Gaussian distribution-like properties. One can also verify that the mean of the distribution increases with increasing background velocities and the increase is higher in the case of delta-correlated noise compared to that for OU noise. It can also be verified that the distribution is more spread out for higher velocities reflecting the corresponding probable longer excursions of the colloid in the same time duration. Additionally, the symmetry of the distribution comes from the fact that the work done by the colloid in the case of constant background flow is a linear function of its trajectory.
\\~\\
On the other hand, in the case of elongational flow where the flow rate ($\dot{\gamma}$) is nonzero, $W_{T}$ becomes a non-linear function of the position of the colloid (see Eq. \ref{eq:work_elongational}). To find $P(W,T)$ in this case, we first calculate the characteristic function, $\mathcal{C}_{W}(\lambda)$, from which we obtain the moments of the distribution following Eq. \ref{eq:moments12}. The value of $\mathcal{C}_{W}(\lambda)$ is then further used in Eq. \ref{eq:work_dist_final} and integrated over $\lambda$ to obtain the exact result for $P(W,T)$. This distribution for the work done in the case of elongational flow field of the fluid is shown in Fig. \ref{fig:work_dist}\rd{b}. Since the shape of this distribution, in contrast to the case for constant flow is no longer symmetric, we also measure its skewness ($\alpha$). Skewness is the parameter that determines the asymmetry of the distribution and can be evaluated as $\alpha = \big[\langle W^3 \rangle - 3\langle W \rangle\sigma^2 - \langle W \rangle ^3\big]/\sigma^3$, where $\langle W^3 \rangle$ is the third moment of the distribution. The distribution is symmetric for $\alpha=0$ and higher the value of $\alpha$, the more asymmetric the distribution becomes. For $\alpha > 0$, the distribution is positive-skewed and for $\alpha < 0$, the distribution is negative-skewed.
\\~\\
The resulting work distribution function of the colloid in the presence of the OU noise is shown in Fig. \ref{fig:work_dist}\rd{b} (filled squares) along with the distribution in the case of white noise (empty squares) for comparison. Distributions were calculated at some arbitrary time $T=1$ and other parameters such as friction coefficient ($\zeta$), stiffness constant ($k$), noise strength ($D$), and the uniform component of the background velocity ($v_0$) were fixed to unity. In the case of the OU noise, the value of $\tau$ was fixed at 0.1. In both the cases (under white noise and OU noise conditions), the distributions are asymmetric which is unlike the case in a constant flow. This asymmetry in the presence of elongational flow appears because of the fact that the work done in this case is a non-linear function of the trajectory of the colloid. This is also evident from Fig. \ref{fig:pdf_position}, where we observed that the distributions become asymmetric in the case of elongational flow, unlike the case for constant flow. A similar effect is also observed in the case of white noise. In this case, the corresponding result is similar to the work distribution of a dumbbell-shaped polymer chain in an elongational flow where the fluctuations were modeled as white noise \cite{sharma2011}. The presence of asymmetry in the work distribution can be described with the help of position distribution which is shown in Fig. \ref{fig:pdf_position}. In the case of constant flow, the particle moves symmetrically in every direction which in turn gives a symmetric work distribution. But, in the case of elongational flow, the particle motion is more likely along the direction of the flow, and hence work done along a particular direction is higher compared to that along the other directions, and the work distribution, therefore, becomes asymmetric.
\begin{figure*}
    \centering
    \includegraphics[width=15cm]{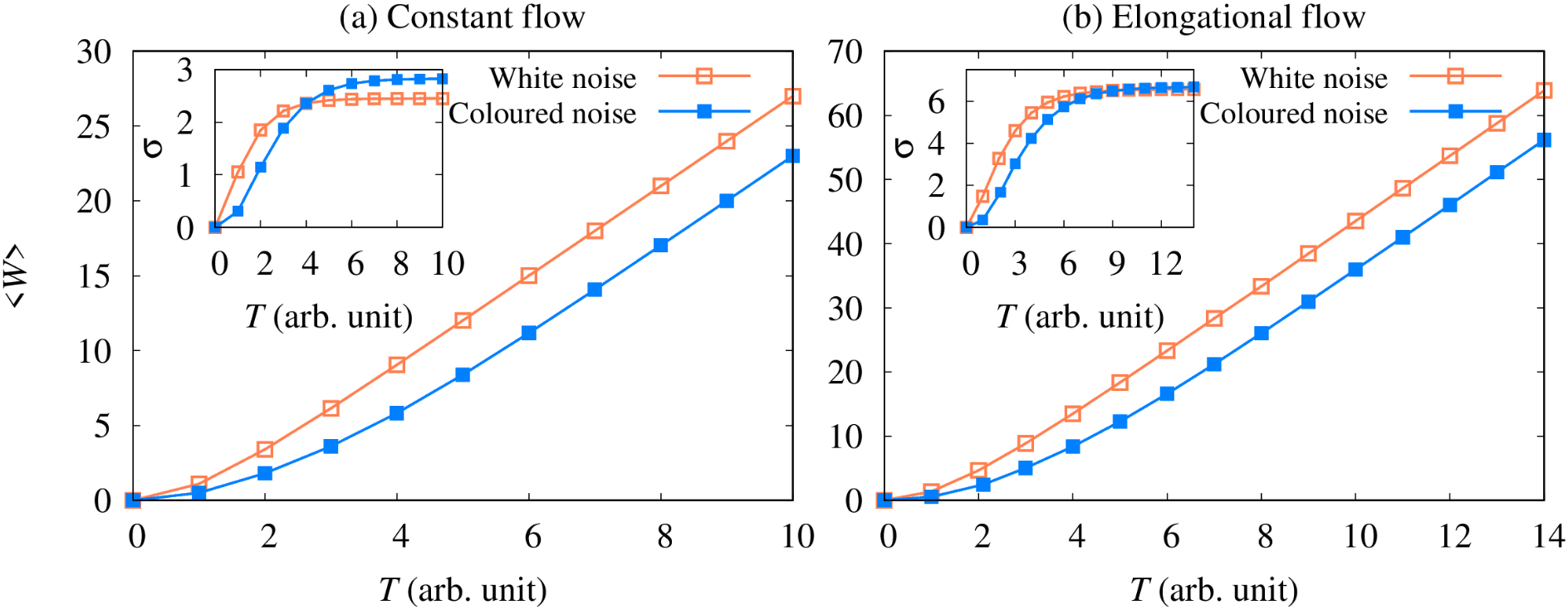}
    \caption{Variation of mean work $\langle W \rangle$ and standard deviation (inset) of $P(W,T)$ as a function of time (arb. unit) for white noise (empty squares) and coloured noise (filled squares) taken in the system for (a) constant flow and (b) elongational flow. In the case of elongational flow, we fixed the flow rate ($\dot{\gamma}$) to 0.3. We set $v_{0}$, $k$, $\tau$, $\zeta$ and $D$ to unity during the calculation.}
    \label{fig:mean_work}
\end{figure*}
\begin{figure}
    \centering
    \includegraphics[width=15cm]{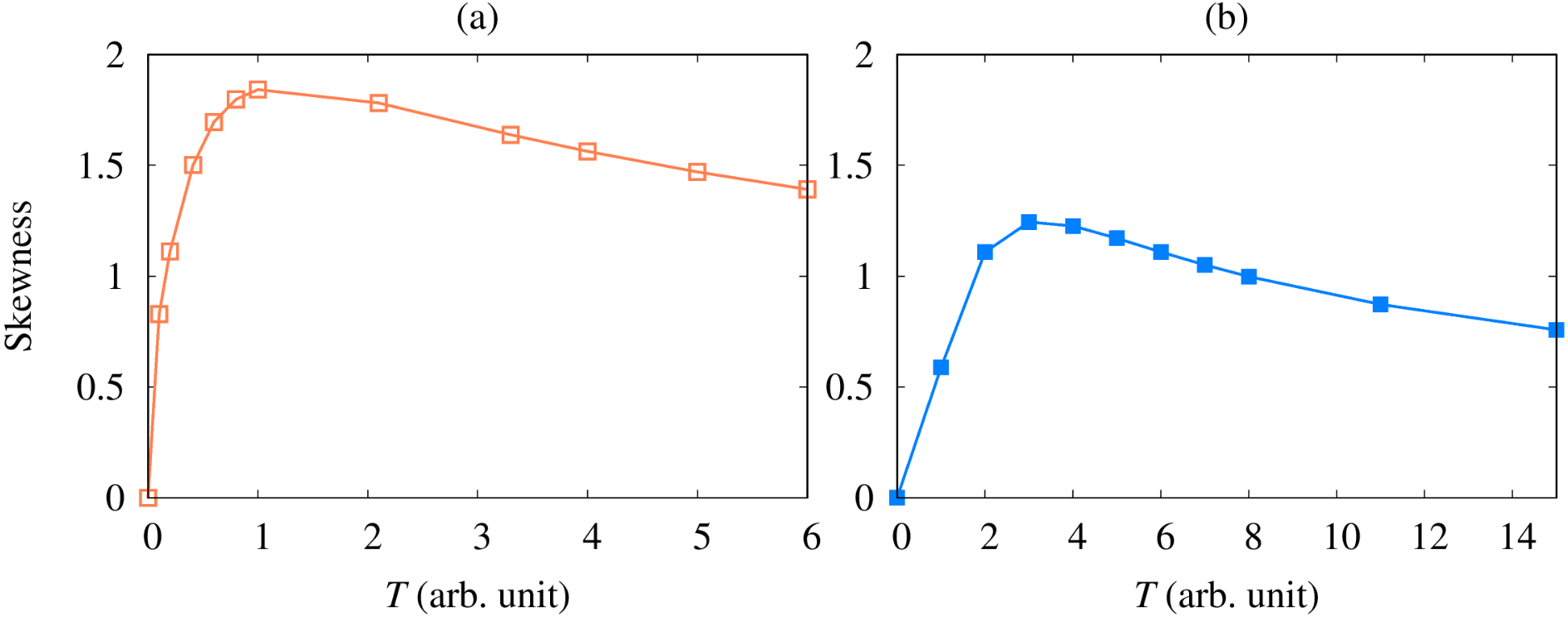}
    \caption{Time evolution of the shape of the work distribution at a constant $\dot{\gamma}=1$ in elongational flow and in the presence of (a) white noise (empty squares), and (b) coloured noise (filled squares). Dots represent the time at which skewness was calculated. We set $v_{0}$, $k$, $\zeta$, $D$ to unity and $\tau$ in the case of OU noise was fixed to 1.}
    \label{fig:skewness}
\end{figure}
\\~\\
The variation of mean work and standard deviation of the work distribution for constant flow and elongational flow are shown in Fig. \ref{fig:mean_work} for both types of noise. The standard deviation ($\sigma$) of the distribution is calculated by taking the square root of the variance. The mean work increases linearly with time for higher times whereas the increase is non-linear in the early time regime for all cases. It should also be noted that the linear increase in the case of white noise appears faster than that for coloured noise for both types of flow. Additionally, the standard deviation for white noise increases in the early time limit whereas it becomes saturated for higher times in both types of flows and for both kinds of noise considered here. This particular behaviour of mean work and standard deviation has a significant role in understanding the fluctuation theorem as well which is discussed later.
\\~\\
Further, the time evolution of the shape parameter (skewness) of work distribution under elongational flow is shown in Fig. \ref{fig:skewness} for both the noise conditions. It shows that the skewness increases very rapidly with time initially and after reaching a maximum, it gradually decreases over time. However, the magnitude of skewness continues to be sufficiently large (greater than zero) even after a significant amount of time has elapsed for both the cases. Nevertheless, as is also evident from Fig. \ref{fig:work_dist}\rd{b}, it is clear that the measure of asymmetry (or the skewness parameter) is higher when white noise is considered in comparison to the case when the system is under the influence of coloured noise.
\begin{figure*}
    \centering
    \includegraphics[width=15cm]{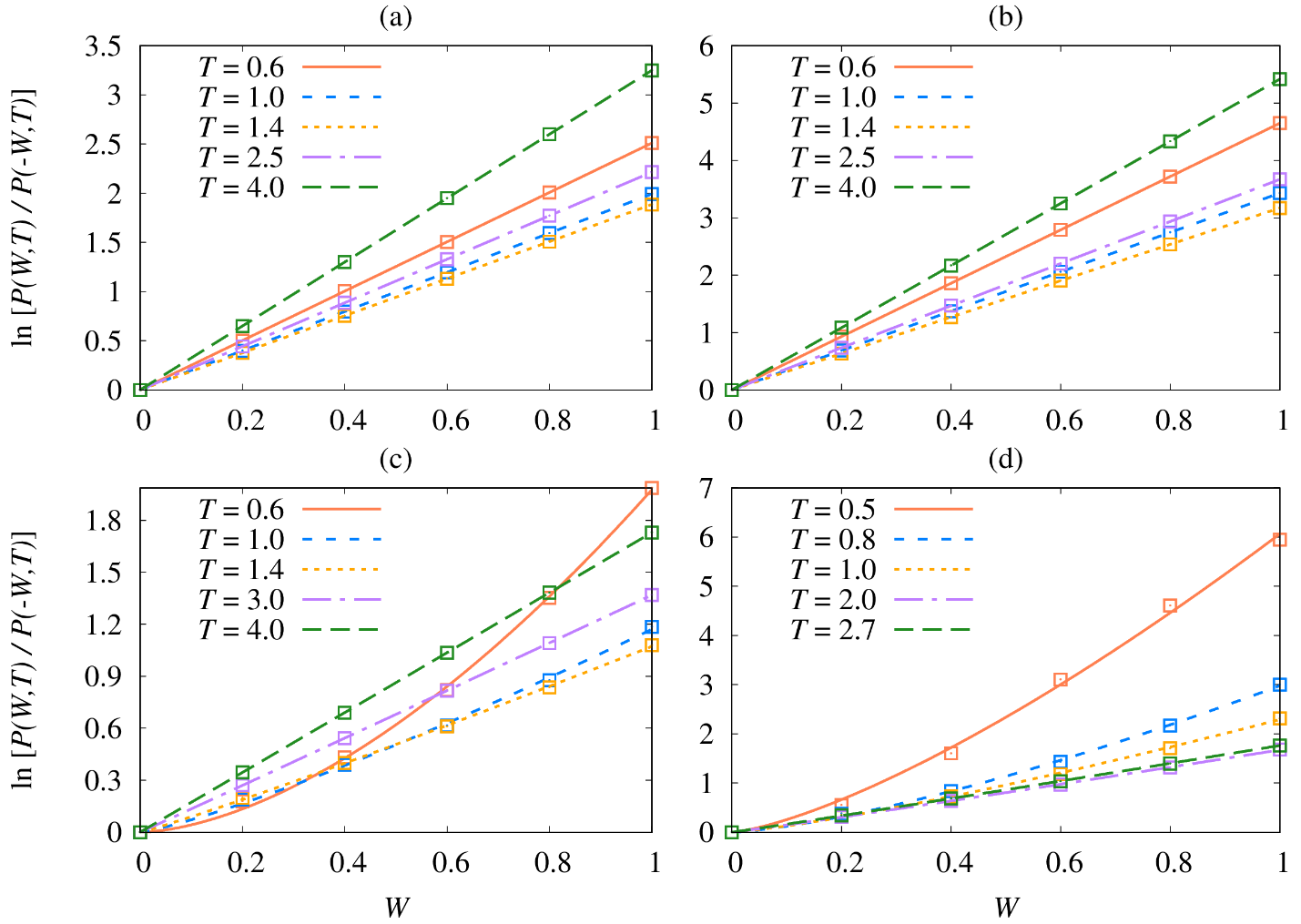}
    \caption{Plot of $\text{ln}[P(W,T)/P(-W,T)]$ versus $W$ ($\beta$ is considered unity) for the case of constant background flow ((a) for white noise and (b) for coloured noise) and elongational flow ((c) for white noise and (d) for coloured noise). In the case of constant background flow, we checked the dependency at various times keeping the background velocity fixed to unity. Similarly, in the case of elongational flow, we checked the dependence for different time limits by fixing the flow rate ($\dot{\gamma}$) to unity. Other parameters ($\zeta$ and $k$) were fixed to unit value and $\tau=0.1$ during the calculations.}
    \label{fig:ft}
\end{figure*}
\\~\\
Next, the function $\ln{[P(W,T)/P(-W,T)]}$ ($=f(W,T)$ let's say) is plotted with respect to $W$ ($\beta$ set to unity) for constant flow and elongational flow in the presence of white and coloured noise to test the validation of the fluctuation theorem. The results are shown in Fig. \ref{fig:ft} in which the curves show different behaviours of the system under different flow properties. In the case of constant flow, $f(W,T)$ for different values of $T$ are straight lines of varying slopes passing through the origin for both white and coloured noise which are shown in Fig. \ref{fig:ft}\rd{a} and Fig. \ref{fig:ft}\rd{b}. The slope of the function initially decreases with increasing value of $T$, but after reaching a minimum, the slope again starts increasing. The slopes themselves depend on the choice of parameter values as well as the amount of time elapsed by the system.
\begin{table}
  \centering
  \renewcommand{\arraystretch}{1.2}
  \begin{tabular}{|p{1cm}|c|c|p{1cm}|c|c|}
    \hline
    \multicolumn{3}{|c|}{\textbf{White noise}} & \multicolumn{3}{|c|}{\textbf{Coloured noise}} \\
    \hline
    $T$ & $a$ & $m$ & $T$ & $a$ & $m$ \\
    \hline
    0.6 & $1.9759\pm0.0236$ & $1.6725\pm0.0447$ & 0.5 & $6.0629\pm0.1188$ & $1.3755\pm0.0618$ \\
    1.0 & $1.1706\pm0.0150$ & $1.2154\pm0.0364$ & 0.8 & $2.9784\pm0.0321$ & $1.3905\pm0.0342$ \\
    1.4 & $1.0716\pm0.0064$ & $1.0814\pm0.0155$ & 1.0 & $2.2900\pm0.0269$ & $1.2544\pm0.0343$ \\
    3.0 & $1.3682\pm0.0010$ & $1.0089\pm0.0018$ & 2.0 & $1.6722\pm0.0067$ & $1.0548\pm0.0102$ \\
    4.0 & $1.7288\pm0.0003$ & $1.0024\pm0.0005$ & 2.7 & $1.7611\pm0.0043$ & $1.0318\pm0.0061$ \\
    \hline
  \end{tabular}
  \caption{Coefficients ($a$) and exponents ($m$) obtained from fitting the curves in Fig. \ref{fig:ft} (c and d) satisfying the fluctuation theorem in the case of elongational flow. The curves were fitted with $f(W) = a W^{m}$ at different $T$ values in the presence of white and coloured noise.}
  \label{tab:table}
\end{table}
\\~\\
Further, Figs. \ref{fig:ft}\rd{c} and \ref{fig:ft}\rd{d}, show the variation of the function $f(W,T)$ at different $T$ values in the case of the elongational flow field and under the influence of white and coloured noise, respectively. Similar to the case of constant flow and as shown in the figure, the slope initially decreases with time and after reaching a minimum it starts increasing. However, unlike the case of constant flow, $f(W,T)$ is a non-linear function of $W$ in the case of the elongational flow field. To further quantify the non-linearity, we have also fitted each curve corresponding to varying $T$ with the function $f(W,T) = a W^{m}$ and the values of $a$ and $m$ for different values of $T$ are given in Table \ref{tab:table}. It is observed that in the presence of white as well as coloured noise, the curve is non-linear for short times (small $T$ values) with $m>1$. However, the value of $m$ decreases and becomes closer to 1 with increasing time resulting in the curves gradually becoming linear. Further, one can see that, the plots follow the trend $f(W,T) \approx W$ for $T \gg \tau$ resembling a phenomenon known as the {\it stationary state fluctuation theorem} (SSFT), expressed as, 
\begin{equation}
    \frac{P(W,T)}{P(-W,T)} \approx \exp{(W)}
\end{equation}
which is found to be valid for large time limits only \cite{van2004, douarche2006}.
\\~\\
Therefore, we see from Fig. \ref{fig:ft} that the work fluctuation in both constant and elongational flow satisfies the fluctuation theorem in the presence of white as well as coloured noise. The non-linear and time-dependent behaviour of FT that we see here for this system was also found earlier in many other systems such as a system in a transient and stationary state in which a harmonically trapped Brownian particle is dragged through a fluid medium \cite{van2004, gomez2011}, a harmonic oscillator in contact with thermostat and under the effect of external force \cite{douarche2006} and a system of a simple electrical circuit consisting of a resistor and capacitor \cite{garnier2005}. In all of the above examples, it was shown that the curves deviate from the $f(W,T) = W$ line and the slopes vary with time and, as the system evolved over a sufficient amount of time, the non-linearity of the curve gradually decreased and approached linearity. Therefore, the work distribution for the colloid in such a system in a non-equilibrium state satisfies the principle of the fluctuation theorem. The fluctuation theorem was also found to be valid in elongational flow for a sufficiently large time where fluctuations were considered as white noise \cite{sharma2011}. In our study, we have reported the work distribution for a colloid and have also established the FT in the presence of OU noise, both for constant background flow and elongational flow.
\begin{figure}
    \centering
    \includegraphics[width=15cm]{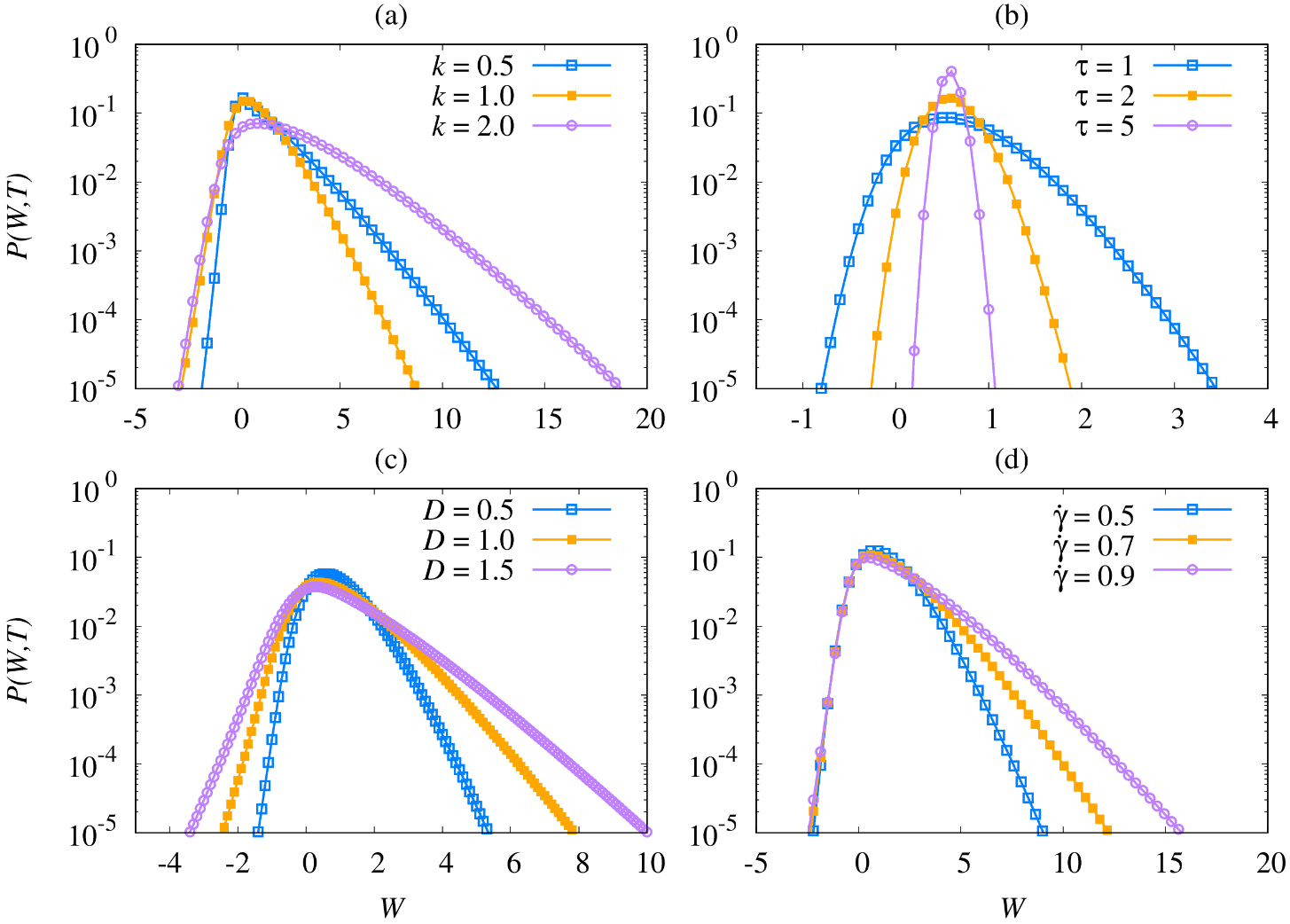}
    \caption{Distributions of work done for the colloid in elongational flow and in the presence of OU noise for varying (a) stiffness constant $k$, (b) relaxation time $\tau$, (c) noise strength $D$, and (d) flow rate $\dot{\gamma}$ (shown with empty circles, empty squares, and filled circles). Distributions were calculated at $T=1$ keeping other parameters except the varying one fixed to unity and $\tau$ fixed to 0.1.}
    \label{fig:parameter_dependence}
\end{figure}
\\~\\
So far, we limited our study to making comparisons of work distributions of a moving colloid in different types of background flows and under different noise conditions. We now look into the dependencies of work distributions of a colloid moving in an elongational flow field and that is influenced by OU noise. Fig. \ref{fig:parameter_dependence} shows how $P(W,T)$ changes with different parameters which were taken to be fixed during earlier computations and how these parameters affect the dynamics of such a colloid. In Fig. \ref{fig:parameter_dependence}\rd{a}, we have shown work distributions for different values of the stiffness coefficient ($k$) of the external harmonic potential. 
As $k$ increases, the peak of the distribution shifts towards higher value of $W$, but the variation of mean and standard deviation with $k$ shows an oscillatory behaviour. Fig. \ref{fig:parameter_dependence}\rd{b} shows distributions for varying relaxation time constants ($\tau$). The standard deviations of the distributions decrease with higher values of $\tau$. This indicates that the distribution gets a short duration of time to spread over for higher relaxation time resulting in lower values of work being sampled across different trajectories. Fig. \ref{fig:parameter_dependence}\rd{c} shows $P(W,T)$ for varying noise strength ($D$). It is evident that the fluctuations in any thermodynamic quantity will increase with increasing noise strength and the result satisfies the argument. We have also shown the variation of $P(W,T)$ with varying flow rates ($\dot{\gamma}$) in Fig. \ref{fig:parameter_dependence}\rd{d}. The mean of the distribution increases with increasing flow rates which suggests that the colloid performs more work in the case of higher flow rates in an elongational flow field.
\\~\\
Until this point, we mainly focused on the behaviour of the colloid in different flow types and different kinds of noise influencing its dynamical properties. We have also studied the temporal evolution of different parameters of the work distribution function and its dependence on the four system parameters. We now look into what this means for the overall thermodynamics of the system. Having computed the work done, the amount of heat exchanged ($Q_T$) with the surroundings up to a time $T$, can be calculated by invoking the first law of thermodynamics i.e., $Q_{T} = W_{T} - \Delta U$, where $\Delta U = k(x_{f}^2+y_{f}^2+z_{f}^2 - x_{0}^2- y_{0}^2- z_{0}^2)/2$ is the change in internal energy of the system. Making use of $W_{T}$ from Eq. \ref{eq:work_elongational}, the amount of heat exchanged is then given by
\begin{equation}
\label{eq:heat}
    Q_{T} = k\int_{0}^{T} \Big[v_{0}(x+y+z) + 2\dot{\gamma}xy \Big] dt - \frac{k}{2}(x_{f}^2+y_{f}^2+z_{f}^2)
\end{equation}
after invoking the initial condition $x_{0} = y_{0} = z_{0} = 0$, as also mentioned earlier. Similar to the work distribution function, the heat distribution function, $P(Q,T)$, representing the probability that $Q_{T}$ amount of heat energy is being exchanged in time $T$ is given by $P(Q,T) = \big\langle \delta \big(Q - Q_{T}\big) \big\rangle$. The characteristic function for the heat exchange can then be obtained following the same technique that was used to compute Eq. \ref{eq:work_dist_fourier} and \ref{eq:characteristic_function} for the calculation of the work distribution. Therefore,
\begin{equation}
\label{eq:char_func_heat}
    \mathcal{C}_{Q}(\lambda) = \big\langle \exp{(- i \lambda Q_{T} )}\big\rangle
\end{equation}
where, $Q_{T}$ can be evaluated from Eq. \ref{eq:heat}. Using this characteristic function, the heat distribution can be calculated as
\begin{equation}
\label{eq:heat_dist}
    P(Q,T) = \int_{-\infty}^{\infty} d\lambda ~\exp{(i\lambda Q)} ~\mathcal{C}_{Q}(\lambda)
\end{equation}
The mean value of the heat energy exchanged with the surrounding medium can also be evaluated from the characteristic function through
\begin{equation}
    \label{eq:mean_heat}
    \big\langle Q \big\rangle = i\frac{\partial}{\partial \lambda} \mathcal{C}_{Q}(\lambda)\bigg|_{\lambda=0}
\end{equation}
Using this set of equations (Eqs. \ref{eq:heat}--\ref{eq:mean_heat}), we have calculated the heat distribution function ($P(Q,T)$) and the mean value of $Q_T$ for a colloid moving in an elongational flow field and influenced by white and coloured noise.
\begin{figure}
    \centering
    \includegraphics[width=15cm]{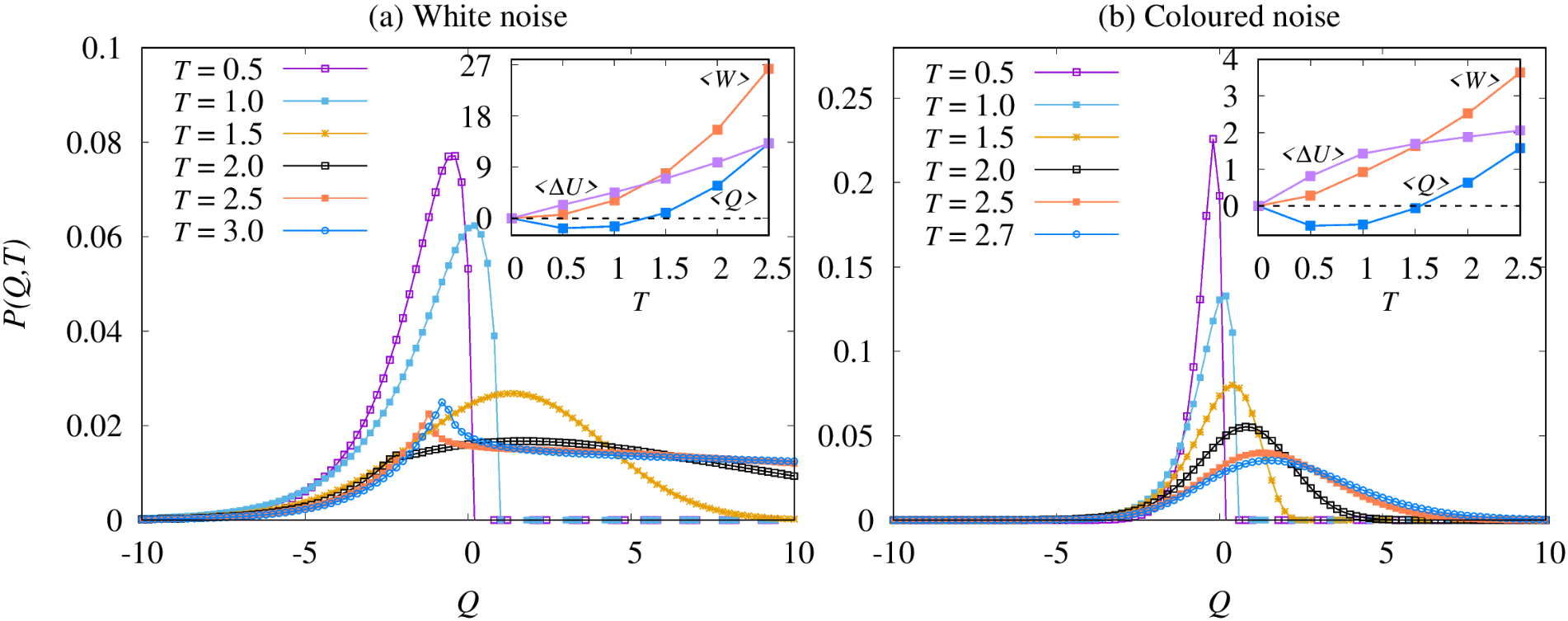}
    \caption{Probability distribution of the heat exchanged $P(Q,T)$ during the motion of the colloid with its surroundings in elongational flow and under the influence of (a) white noise and (b) coloured noise at different final times. The figures in the inset show the change of $\langle Q \rangle$, $\langle W \rangle$ and $\langle \Delta U \rangle$ with time. Horizontal dotted lines in the inset of each figure show $\langle Q \rangle = 0$ line. Calculations were done keeping $\zeta$, $k$, $\dot{\gamma}$, $D$, $v_{0}$ fixed to unity, and $\tau$ in the case of coloured noise is fixed to 0.1.}
    \label{fig:heat_dist}
\end{figure}
\\~\\
The final results for the heat distribution of the colloid moving in the elongational flow field and under the influence of the two different noise types are shown in Fig. \ref{fig:heat_dist}. One can see that at a sufficiently small time ($T=0.5$), the probability of positive heat exchange is approximately zero, but it gradually increases with time. Analogous to the distribution, the mean value of the heat exchange (shown in the inset of Fig. \ref{fig:heat_dist}) is also negative at short times. It indicates that the system consumes heat energy from the surroundings to perform the work required to move the colloid from one position to another. A similar phenomenon of heat consumption in the small time limit was earlier also found in a system having multiple coupled harmonic oscillators kept at different temperatures \cite{agarwalla2014} and a system with a trapped Brownian oscillator in aging gelatin droplet \cite{gomez2012}. After a sufficient time has elapsed ($T > 1.5$), the mean heat exchange becomes positive and the system starts dissipating heat into the surrounding medium. At the time at which this transition occurs, the distribution, $P(Q,T)$, becomes continuous with significant probability for positive $Q$ values as well. The high probability for positive heat exchange continues beyond the transition time but the probability distribution itself starts to exhibit a singularity in the case of white noise (see Fig. \ref{fig:heat_dist}\rd{a}). On the other hand, in the case of coloured noise, the heat distribution shows a similar transition, but no discontinuity or singularity is observed for higher $T$ values. This kind of transition in the heat distribution was also found when the fluctuation phenomena was studied in an electrical circuit having a resistor and a capacitor in parallel \cite{garnier2005}. Heat distributions are known to generally have a discontinuity in these kinds of systems in which a passive colloid is driven by an external harmonic potential and noise \cite{garnier2005, chatterjee2010, imparato2007, gomez2011, berut2016, manikandan2017}. Additionally, the spread of the heat distribution is smaller in the case of coloured noise compared to that in white noise which is consistent with the results obtained for the work distribution and the conditional probability distribution for the final position of the colloid as well, as shown in Figs. \ref{fig:pdf_position} and \ref{fig:work_dist}, respectively.
\\~\\
Now let us discuss the overall thermodynamic picture of the system in terms of the fluctuation theorem, work done by the colloid, and the heat exchanged with the surrounding medium. During the study of the fluctuation theorem we noticed that at a small time limit, the probability of positive work is much higher compared to that of the negative work. The ratio initially decreases with time and after reaching a minimum, it starts increasing again. This particular phenomenon can be explained with the help of Fig. \ref{fig:mean_work}, in which the temporal evolution of the mean and standard deviation of the work distribution is shown. When $T$ is very small, the ratio $P(W,T)/P(-W,T)$ is very large, but as $T$ increases, the standard deviation of $P(W,T)$ also increases which implies that the width of the distribution becomes broader with time. As a result, the probability of negative work also increases with time. After sufficient time has elapsed (in both noise types), the standard deviation becomes saturated and the mean of the distribution continues to increase linearly with time. This basically implies that the distribution shifts towards the higher $W$ without getting broadened in width and because of this, the probability of positive work increases and that of negative work decreases with $T$. This explains why the ratio again starts increasing after a certain amount of time and continues to increase monotonically. Although the slope of the ratio was previously found to oscillate with time \cite{douarche2006}, it is not the same for our system. The insets of Fig. \ref{fig:heat_dist} show that at early times, the system consumes heat from the surroundings and part of the heat is used to increase the internal energy, and the remaining amount is used to perform some work which is required for the system to move from one state to another. The entropy change of the medium $\Delta S_{m} = Q/T^{\prime}$ also becomes negative. After some time, the system stops consuming heat, in fact, it starts dissipating heat to the surroundings, and therefore the entropy of the medium increases. Further, consistent with the results of Fig. \ref{fig:pdf_position} and Fig. \ref{fig:mean_work}, the average work done and the average change in internal energy increases with time as it is dependent on the amount of displacement of the colloid from the initial position.
\section{Conclusions}
\label{sec:conclusions}
In summary, we deduce that the work distribution of the colloid is symmetric when it moves in a constant flow from the initial position to the final position during a time interval of $T$ irrespective of whether it is under the influence of either delta-correlated noise or OU noise. But, in the case of elongational flow, the work distribution becomes asymmetric for both types of noise. This is a direct effect of the non-linearity introduced in the work done as a result of the elongational flow field. We have also studied the temporal dynamics of different distribution parameters, such as mean work and standard deviation. Specifically, in the early time limit, the increase in mean work is nonlinear in time, whereas it becomes linear after a finite amount of time for each type of flow and under both noise types. We also observed that the linear increase of mean work appears quicker in delta-correlated noise compared to that in OU noise for each flow. Additionally, the standard deviation of each distribution becomes saturated after a finite amount of time. The fluctuation theorem is also satisfied for the system considered in this study. We have also looked at how the work distribution changes with varying system parameters, such as stiffness constant $k$, relaxation time $\tau$, noise strength $D$, and flow rate $\dot{\gamma}$. Out of these four system parameters, the work distribution narrows only for increasing $\tau$ values, while it, in general, broadens for increasing values of other system parameters. This is because of the correlated nature of the fluctuations, which do not allow the colloid to undergo large excursions. This phenomenon is then also reflected in the work distribution function. Similarly, a study of other thermodynamic variables also gives valuable information about the system. At the small time limit, heat is consumed by the system to perform work. However, in the long time limit, as the system performs more work as a result of the increased displacement of the colloidal particle due to the elongational flow field, heat is dissipated into the surroundings.
Therefore, the present study provides an insight into a yet unexplored aspect of the dynamics of a colloidal particle in a flow field, specifically, while it is also under the influence of a type of correlated noise. Future extensions of this study looking into the effects of other kinds of noise will really open up avenues for better theoretical understanding and eventually greater control over experimental studies in real biological systems as well.
\appendix
\section{Calculation of the Jacobian}
\label{app:jacobian}
\noindent
The equation of motion of a colloidal particle moving in a medium and under the influence of external harmonic potential can be expressed by an overdamped Langevin equation
\begin{equation}
\label{eq:jacobian_langevin}
    \zeta \dot{\vec{r}} - \zeta \vec{v}(\vec{r}) + \frac{\partial U(\vec{r})}{\partial\vec{r}} = \vec{\eta}(t)
\end{equation}
where, $\vec{v}(\vec{r}) = \vec{v}_{0} + \dot{\gamma}\kappa\vec{r}$ is the velocity of the background medium exhibiting elongational flow. For the case of the constant flow, $\dot{\gamma}=0$.
\\~\\
To calculate the Jacobian of the coordinate transformation ($J[x,y,z]$ as mentioned in Eq. \ref{eq:prob_dist_positions}) from $\vec{\eta}$ to $\vec{r}$ \cite{sharma2011, chatterjee2010}, we modified Eq. \ref{eq:jacobian_langevin} as
\begin{equation}
\label{eq:jacobian_langevin_modified}
    \zeta\dot{\vec{r}}-\zeta\vec{v}_{0}-\vec{D}\cdot\vec{r} = \vec{\eta}(t)
\end{equation}
where $\vec{D}=-k\vec{I}$ for constant flow and $\vec{D}=\zeta\dot{\gamma}\kappa-k\vec{I}$ for elongational flow, $\vec{I}$ is the unit tensor. We can then write Eq. \ref{eq:jacobian_langevin_modified} in discrete form which reduces to
\begin{equation}
\label{eq:jacobian_langevin_discretized}
    \vec{\eta}(t_{i})=\zeta\frac{\vec{r}(t_{i})-\vec{r}(t_{i-1})}{\Delta t} - \vec{D}\cdot\frac{\vec{r}(t_{i})+\vec{r}(t_{i-1})}{2} - \zeta \vec{v}_{0}
\end{equation}
where $i=1,2,...,N$, corresponds to different time steps. $J[x,y,z]$ can be found by calculating det$\big[\partial \vec{\eta}(t_{i})/\partial \vec{r}(t_{j})\big]$, $i,j=1,2,...,N$, which eventually gives a $N \times N$ lower triangular matrix. The determinant can then be easily calculated as follows
\begin{equation}
\label{eq:jacobian}
\begin{split}
    J[x,y,z] &= \prod_{i=1}^{N} \Big(\zeta/\Delta t - \vec{D}\cdot\vec{I}/2 \Big) = \prod_{i=1}^{N} \Big(\zeta/\Delta t + k/2\Big)^{3} \\
    &=(\zeta/\Delta t)^{3N}\prod_{i=1}^{N}\Big(1+3k\Delta t/2\zeta +\mathcal{O}(\Delta t^{2})\Big) \\
    &=(\zeta/\Delta t)^{3N}\exp(3k\Delta t/2\zeta)
\end{split}
\end{equation}
The total time ranging from 0 to $T$ is divided into $N$ equal segments of equal width $\Delta t$ such that the time elapsed after $i^{\text{th}}$ step is given as $t_{i}=i\Delta t$. For the continuum limit $N\rightarrow\infty$, $\Delta t\rightarrow 0$ and $N\Delta t\rightarrow T$, the above equation reduces to $J[x,y,z]=(\zeta/\Delta t)^{3N}\exp(3kT/2\zeta)$. 

\section{Conditional PDF for the case of Gaussian white noise}
\label{app:pdf_delta}

\noindent
The Gaussian white noise usually accounting for thermal fluctuations considered to be produced from random collisions of the colloid with other surrounding particles, has the following properties:
\begin{subequations}
\label{eq:white_noise}
    \begin{equation}
        \big\langle \vec{\eta}_{i}(t) \big\rangle = 0
    \end{equation}
    \begin{equation}
        \big\langle \vec{\eta}_{i}(t) \vec{\eta}_{j}(t^{\prime}) \big\rangle = 2\zeta k_{\text{B}}T^{\prime}\delta_{ij} \delta (t-t^{\prime})
    \end{equation}
\end{subequations}
Since the white noise is Gaussian distributed, the probability distribution of the noise can be written as,
\begin{equation}
\label{eq:prob_dist_white}
    P[\vec{\eta}] \propto \exp\bigg\{-\frac{1}{8\zeta k_{\text{B}}T^{\prime}}\int_{0}^{T}dt\vec{\eta}(t)^{T}\vec{\eta}(t)\bigg\}
\end{equation}
Substituting the value of $\vec{\eta}(t)$ from Eq. \ref{eq:langevin} into Eq. \ref{eq:prob_dist_white}, the probability can be evaluated as
\begin{equation}
\begin{split}
\label{eq:prob_dist_position_delta}
    P[x,y,z] \propto J[x,y,z] \exp\bigg\{-\frac{1}{8\zeta k_{\text B}T^{\prime}} \int_{0}^{T} \Big[ \zeta^2(\dot{x}^2+\dot{y}^2+\dot{z}^2) - 2\zeta^{2}v_{0}(\dot{x}+\dot{y}+\dot{z}) \\ 
    + 2\zeta k(x\dot{x}+y\dot{y}+z\dot{z}) - 2\zeta^{2}\dot{\gamma}(\dot{x}y+x\dot{y}) - 4\zeta k\dot{\gamma}xy + (k^2+\zeta^{2}\dot{\gamma}^2)(x^2+y^2) + k^{2}z^{2} \\
    + (2\zeta^{2}\dot{\gamma}v_{0}-2\zeta kv_{0})(x+y) 
    - 2\zeta kv_{0}z + 3\zeta^{2}v_{0}^{2} \Big] \bigg\}
\end{split}
\end{equation}
$J[x,y,z]$ is the Jacobian for transforming the coordinates from $\vec{\eta}$ to $\vec{r}$ whose calculation is given in Appendix \ref{app:jacobian}.
The conditional probability distribution for finding the colloid at ($x_{f},y_{f},z_{f}$) after a finite time $T$ given that it was at ($x_{0},y_{0},z_{0}$) at $t=0$ is given by
\begin{equation}
\begin{split}
\label{eq:cond_prob_delta}
    P(x_{f},y_{f},z_{f},T | x_{0},y_{0},z_{0}) & \propto e^{3kT/2\zeta}~e^{-k(x_{f}^{2}+y_{f}^{2}+z_{f}^{2}-x_{0}^{2}-y_{0}^{2}-z_{0}^{2})/4k_{\text B}T^{\prime}} \\
    & \times\int_{x(0)=x_{0}}^{x(T)=x_{f}} \mathcal{D}[x] \int_{y(0)=y_{0}}^{y(T)=y_{f}} \mathcal{D}[y] \int_{z(0)=z_{0}}^{z(T)=z_{f}} \mathcal{D}[z] ~e^{-S[x,y,z]}
\end{split}
\end{equation}
where $\mathcal{D}[x]$, $\mathcal{D}[y]$ and $\mathcal{D}[z]$ represent the path integrals over $x$, $y$ and $z$ between the end points ($x_{0},y_{0},z_{0}$) and ($x_{f},y_{f},z_{f}$) and $S[x,y,z]$ represents the action, defined as
\begin{equation}
\label{eq:action_delta}
    S[x,y,z] = \int_{0}^{T} dt ~\mathcal{L}(x,y,z,\dot{x},\dot{y},\dot{z},t)
\end{equation}
Here, $\mathcal{L}$ is the Lagrangian of the system given by
\begin{equation}
\begin{split}
\label{eq:lagrangian_delta}
    &\mathcal{L} (x,y,z,\dot{x},\dot{y},\dot{z},t) = \frac{1}{8\zeta k_{\text B}T^{\prime}}\Big[ \zeta^2(\dot{x}^2+\dot{y}^2+\dot{z}^2) - 2\zeta^{2}v_{0}(\dot{x}+\dot{y}+\dot{z}) - 2\zeta^{2}\dot{\gamma}(\dot{x}y+x\dot{y}) \\
    &- 4\zeta k\dot{\gamma}xy + (k^2+\zeta^{2}\dot{\gamma}^{2})(x^2+y^2) + k^{2}z^{2} + (2\zeta^{2}\dot{\gamma}v_{0} - 2\zeta kv_{0})(x+y) - 2\zeta kv_{0}z + 3\zeta^{2}v_{0}^{2} \Big]
\end{split}
\end{equation}
\\~\\
The most probable trajectory of the colloid between two given endpoints in such a system can be obtained by using the Euler-Lagrange equation of motion, which is given by
\begin{equation}
\label{eq:euler_lagrange_eom_delta}
    \frac{\partial\mathcal{L}}{\partial r_{i}} - \frac{d}{dt}\bigg(\frac{\partial\mathcal{L}}{\partial \dot{r}_{i}}\bigg)=0
\end{equation}
Using the Lagrangian in Eq. \ref{eq:euler_lagrange_eom_delta}, the equations of motion of the colloid along individual components can be obtained as
\begin{equation}
\label{eq:eom_delta}
    \ddot{\vec{r}} + R\vec{r} + S\vec{I} = 0
\end{equation}
where $R=\Big(\begin{smallmatrix}
    -\alpha_{1} & \alpha_{2} & 0 \\
    \alpha_{2} & -\alpha_{1} & 0 \\
    0 & 0 & -\beta_{1}
\end{smallmatrix}\Big)$, $S=\Big(\begin{smallmatrix}
    \alpha_{3} \\
    \alpha_{3} \\
    -\beta{2}
\end{smallmatrix}\Big)$, $\vec{I}$ is the $3\times3$ identity matrix,
$\alpha_{1} = \frac{k^2}{\zeta^2} + \dot{\gamma}^{2}$, $\alpha_{2} = \frac{2k\dot{\gamma}}{\zeta}$, $\alpha_{3} = \dot{\gamma}v_{0} - \frac{kv_{0}}{\zeta}$ and $\beta_{1} = \frac{k^2}{\zeta^2}$, $\beta_{2} = \frac{kv_{0}}{\zeta}$. Solutions to the above equations can be obtained by integrating them using the boundary conditions $x(0)=x_{0}$, $x(T)=x_{f}$, $y(0)=y_{0}$, $y(T)=x_{f}$, $z(0)=z_{0}$, $z(T)=z_{f}$. Similar to the case with OU noise, we solved Eq. \ref{eq:eom_delta} numerically by setting $x_{0}=y_{0}=z_{0}=0$ and the total time taken to be unity. Other parameters were set to have the same values as before. The solutions were then used to evaluate the action using Eq. \ref{eq:action_delta} from which we calculated the normalized probability distribution for the final position of the colloid, given by 
\begin{equation}
\label{eq:final_pdf_delta}
\begin{split}
    P_{N}(x_{f},y_{f},z_{f},T | x_{0},y_{0},z_{0}) = A_{1} ~\exp\Big[ A_{2}(x_{f}+y_{f}) + A_{3}x_{f}y_{f}+ A_{4}(x_{f}^2+y_{f}^2) +A_{5} z_{f} +A_{6} z_{f}^2 \Big]
\end{split}
\end{equation}
where $A_{i}$'s are some constants that depend on a particular set of parameter values. The flow rate is taken to be unity for the calculation of the PDF. The distribution is shown in Fig. \ref{fig:pdf_position} both for white noise (left) and coloured noise (right) at two different times.

\bibliographystyle{unsrt}
\bibliography{refs}

\end{document}